\documentclass[12pt,preprint]{aastex} 
\usepackage{epsfig,lscape}

\begin{document}

\title{Constraints on Velocity Dispersion Function of Early-type Galaxies
from the Statistics of Strong Gravitational Lensing}

\author{Kyu-Hyun Chae}
\affil{Department of Astronomy and Space Sciences, Sejong University,
  98 Gunja-dong, Gwangjin-Gu, Seoul 143-747, Republic of Korea}

\begin{abstract} 
We use the distribution of gravitationally-lensed image separations observed
in the Cosmic Lens All-Sky Survey (CLASS) and the PMN-NVSS Extragalactic Lens 
Survey (PANELS), which are (nearly) complete for the image separation range 
$0''.3 \le \Delta \theta \le 6''$, to constrain a model velocity dispersion
function (VF) of early-type galaxies. Assuming a current concordance 
cosmological model and adopting a singular isothermal ellipsoid (SIE) model for
galactic potentials, we consider constraining both a characteristic velocity 
dispersion (parameter $\sigma_*$) and the shape of the function 
(parameters $\alpha$ and $\beta$; Sheth et al.\ 2003) 
for $0.3 \la z \la 1 $. If all three parameters are allowed to vary, 
then none of the parameters can be tightly constrained
by the lensing data because of the small size of the sample. If we fix 
the shape of the function by either the SDSS local stellar VF or an inferred
local stellar VF based on the SSRS2 galaxy sample, then the constrained values
of $\sigma_*$ are nearly equal to the corresponding stellar values; we have 
$f_{\rm SIE/center} (\equiv \sigma_{*{\rm SIE}}/\sigma_{*{\rm center}})= 
0.90\pm 0.18 {\rm (SDSS)}$ or $1.04 \pm 0.19 {\rm (SSRS2)}$ assuming 
non-evolution of the function between the present epoch and $z \sim 0.6$. 
Finally, using only the CLASS statistical sample (Browne et al.\ 2003) and 
thus including an absolute multiple-imaging probability, 
we find that the SDSS stellar VF may have significantly underestimated 
the abundance of morphologically early-type galaxies.
\end{abstract}


\section{Introduction} 
One of the fundamental parameters for an early-type galaxy is  the 
(1-dimensional) velocity dispersion, i.e.\ the standard deviation of 
the line-of-sight velocities of objects. 
In general, the velocity dispersion is anisotropic and varies as a function of
projected radius from the galactic center although it is expected to be
constant for certain isothermal galaxy models. Hence, for a given galaxy
there arise different velocity dispersions depending on the probed scales. 
For example, the velocity dispersion determined from 
the spectroscopic observation of a galaxy within a given 
aperture refers to the stellar kinematics in an inner region (usually, the
central few kpc region) of the galaxy, while the velocity dispersion 
inferred from the extended X-ray emitting gas temperatures refers to the 
kinematics of (the galaxy's mass dominating) halo. 

The distributions of (suitably chosen) velocity dispersions among all 
early-type galaxies at a given cosmic epoch, or the velocity dispersion 
functions (VFs) of early-type galaxies, are crucial observables that can 
provide powerful constraints on models of galaxy formation and evolution 
(see, e.g., Kochanek 2001; Sheth et al.\ 2003). 
Specifically, VFs for different scales 
(e.g., the central region, the effective radius region, the extended halo 
region) can be compared with one another as well as predictions of models 
of galaxy formation and evolution. Conventionally, central stellar VFs of 
early-type galaxies have been inferred from early-type luminosity 
functions via an adopted power-law relation between luminosity and velocity 
dispersion (i.e.\ the `Faber-Jackson' relation) $\sigma \propto L^{1/\gamma}$
(e.g., Shimasaku 1993; Gonzalez et al.\ 2000; Chae 2003).  

Recently, Sheth et al.\ (2003) have pointed out that the spread in a
Faber-Jackson relation results in a systematic bias for the inferred central 
stellar VF. Sheth et al.\ (2003) have then directly derived a central stellar 
VF based on Sloan Digital Sky Survey (SDSS) spectroscopic data of $\sim 9000$
early-type galaxies (Bernardi et al.\ 2003). Mitchell et al.\ (2005) have 
updated (the normalization of) the SDSS VF using a larger galaxy sample.
Bernardi et al.\ (2003) and Mitchell et al.\ (2005) have applied 
both surface brightness profile and
spectroscopic criteria to SDSS galaxies to classify them; in particular, a 
galaxy was assigned early-type only when it satisfied all the applied 
criteria. As discussed in Chae (2003), accurately classifying large numbers 
galaxies is the major difficulty in deriving reliable type-specific 
luminosity functions; the same is true for deriving reliable velocity 
dispersion functions. The conventional method of classifying galaxies by the 
visual inspection of their morphological appearances is probably the most 
reliable method (see, e.g., Marzke et al.\ 1998; Chae 2003; Madgwick 2003) 
but unfortunately cannot be applied to very large samples of galaxies 
such as the SDSS sample. In this respect, the SDSS VF should be taken with 
some caution until it is verified that the galaxy classification process
(Bernardi et al.\ 2003; Mitchell et al.\ 2005) does not suffer from any 
significant systematic bias.\footnote{It is interesting to note that Bernardi
et al.\ (2003) assign $\approx$ 14\% of their galaxies early-type while 
Marzke et al.\ (1998), who classified by morphological appearances, 
assign $\approx$ 30\% early-type and Madgwick et al.\ (2002), 
who classified 2dFGRS galaxies by spectroscopic criteria, 
assign $\approx$ 40\% early-type.}

In this work, we consider constraining a model VF of early-type 
galaxies using the statistics of strong gravitational lensing.
We use mainly the  distribution of image separations to constrain the
shape of the VF and the characteristic velocity dispersion $\sigma_*$.
For this purpose we need a sample of lensed systems 
that are complete for image separations (but may not be 
complete otherwise; see \S 2). The best such sample is provided by the 
lensed systems that have been discovered in the Cosmic Lens All-Sky Survey 
(CLASS; Myers et al.\ 2003; Browne et al.\ 2003) of radio-loud sources and 
the PMN-NVSS Extragalactic Lens Survey (PANELS; Winn et al.\ 2001b), which
is a southern sky survey somewhat similar to the CLASS. These surveys 
have discovered a total of 26 lensed systems which form a (nearly) complete
sample for the image separation range $0''.3 \le \Delta \theta \le 6''$.
We also consider constraining the normalization of the VF (which is 
proportional to the abundance of early-type galaxies). We cannot use 
the whole sample of lensed radio sources mentioned above for this purpose. 
This is because not all of them pertain to a radio source sample that
can satisfy well-defined observational selection criteria and thus provide
an absolute probability of multiple-imaging. 
The largest sample of radio sources satisfying well-defined observational 
selection criteria is the CLASS statistical sample of 8958 sources containing
13 multiply-imaged sources (Browne et al.\ 2003; Chae 2003). 
We use this sample to constrain the normalization of the VF.

Since the time it was pointed out (e.g., Fukugita, Futamase, \& Kasai 1990; 
Turner 1990) that lensing rate increases rapidly with increasing $\Lambda$,
strong lensing statistics have been most often used to constrain cosmological
parameters under the assumption that adopted VFs were correct (e.g.\ 
Kochanek 1996; Helbig et al.\ 1999; Chae et al.\ 2002, 2004). 
In this work, we use lensing statistics, specifically image separation
statistics, for a different purpose;
namely, we constrain a model VF under the assumption that the adopted
cosmological model is correct. Here it is worth making two points. First,
a sufficiently large sample of lenses can constrain simultaneously both 
cosmological parameters and a VF but the current lens sample is not so large
that we are forced to make some assumptions. Second, given the widespread 
convergence of cosmological measurements toward the current concordance
cosmological model (e.g., Riess et al.\ 1998; Perlmutter et al.\ 1999;
 Spergel et al.\ 2003), namely the 
spatially-flat $\Omega_{\rm m} \approx 0.3$ universe, it is at least as 
interesting to use lensing statistics for deriving information on galaxies as
for constraining cosmological parameters. 

Constraining a VF through image separation distributions of 
gravitationally-lensed systems is unique and worthwhile in the following 
respects. First, the image separation of a gravitationally-lensed system 
is dependent on both luminous mass and dark mass within the Einstein radius
of the lensing galaxy, which is of order of its effective radius. 
Consequently, the image separation probes the scale that is intermediate
compared with the scales probed by aperture-limited spectroscopic observations
and X-ray observations and encodes the information on the dynamics of
both luminous mass and dark mass in the galaxy. 
Second, the angular sizes of image splitting in lensed 
systems can be measured accurately, particularly through radio observations 
as it is the case for the CLASS and the PANELS lenses.
However, some care should be taken in turning an observed image separation 
distribution into a VF for the following reasons. First, an image-splitting 
size can be translated into a velocity dispersion 
only through a galaxy model (see \S 3 for further).
For example, a single-component isothermal galaxy model can be adopted and 
then the inferred velocity dispersion would be some sort of effective quantity
for luminous and dark mass components.\footnote{In the literature including
the author's own, the velocity dispersion derived from an image
separation has often been mistakenly referred to as the 
``dark matter velocity dispersion'' (see  \S 3 for further).} 
Second, about a quarter of the multiply-imaged systems
involve multiple galaxies within their Einstein radii, so that the
image separations in those systems do not correspond to the properties
of single galaxies.

The lensing galaxies we use in this work lie within the redshift range of
$0.3 \la z  \la 1$. This means that we are effectively constraining a VF 
at $z \sim 0.6$. This also means that we can constrain some aspects of 
galaxy evolution using
the lens sample.\footnote{Such an analysis has been done in Chae \& Mao
(2003) using an inferred local stellar VF. See, also, Ofek, Rix, \& Maoz
(2003).} However, the focus of this work is somewhat different; we obtain
a solely lensing-based VF for $z \sim 0.6$ and then compare it with (nearly)
local stellar VFs assuming only a passive evolution of early-type galaxies 
between the two epochs.

Below in \S 2 we briefly describe the CLASS data and the methodology. In \S 3 
we present the constraints on a model velocity dispersion function of
early-type galaxies and discuss their implications.

\section{Data and Method}

The CLASS\footnote{Throughout this paper CLASS is meant to include its
predecessor survey, the Jodrell Bank-VLA Astrometric Survey 
(King et al.\ 1999).} has observed a total of 16,503 targeted 
flat-spectrum radio sources resulting in the 
discovery of 22 cases of multiple-imaging with image separation 
$\Delta\theta \ge 0''.3$ (Myers et al.\ 2003; Browne et al.\ 2003).  
The total sample may not be suitable for
multiple-imaging rate statistics because not all sources satisfy
a set of source selection criteria.\footnote{Only a subsample of 8958 sources
containing 13 multiply-imaged sources satisfy well-defined observational
selection criteria and is referred to the CLASS statistical sample.
This sample has been extensively analyzed for the purpose of deriving
information on cosmological parameters and galaxy properties (e.g.,
Chae 2003; Chae et al.\ 2002, 2004; Chae \& Mao 2003).}
  For example, a fraction of the sample
does not strictly satisfy the flat-spectrum criterion, so that their parent
population may be different from that of flat-spectrum sources. This means
that the magnification bias and redshift distribution estimated for 
flat-spectrum sources could not be applied to the non-flat-spectrum sources.
However, as long as only image separation statistics is concerned,
source population properties do not matter because image separations are
determined by the properties of the lensing galaxies. In other words,
whatever the source may be, for the same lensing geometry, lensing galaxy, and
cosmology the image separation must be the same. The only requirement of the
sample for image separation statistics is that the lens discovery process 
does not involve any particular bias against certain image separation range.
The entire CLASS sample satisfies this requirement for 
$0''.3 \le \Delta\theta \le 6$. 
The PANELS radio survey of partial southern sky 
($0\deg > \delta > -40\deg $) using the same methodology of the CLASS has
observed 4097 flat-spectrum sources and discovered 4 multiple-imaging
cases to date (Winn et al.\ 2000, 2001a, 2002a,b). The lens search
process for this survey has not yet been completed and so the lens sample
is not quite complete. However, since the lens search process is not biased
against any particular image separation range, we could add the PANELS
sample to the CLASS sample for our image separation statistics.

The properties of the 22 multiply-imaged CLASS systems are summarized in 
Browne et al.\ (2003). The properties of the 4 PANELS multiply-imaged 
systems can be found Winn et al.\ (2000, 2001a, 2002a,b). In Table~1 we 
summarize the essential properties of the 26 multiply-imaged systems.
Not all of these systems can be used for our analysis. 
First, we exclude the systems whose lensing galaxies are spiral galaxies 
(Group C). Second, we exclude the systems that contain multiple galaxies 
within their critical curves, since the observed image separations for 
the systems do not correspond to the properties of single galaxies (Group~D).
Notice particularly that the secondary (and tertiary) galaxies in these 
systems are comparable in mass to the corresponding primary galaxies; see
\S 3.1 of Chae~(2003) for details. This means that both the convergences and
shears of the secondary (and tertiary) galaxies are significant for the 
the systems of Group~D.\footnote{In other words, the secondary (and tertiary)
galaxies cannot simply be regarded as perturbation terms to the main lensing 
potentials.} Notice also that Group~D does not include lens systems that
contain proximate galaxies outside the critical curves (e.g.\ 0414+054,
1030+074, 1152+199), since those nearby galaxies can normally be accounted for
by external shears and are not likely to contribute significantly to the 
observed image separations (see below however).
Third, we exclude any systems whose lensing scenario details are so 
uncertain that we cannot reliably interpret their image separations (Group~E).
Group~E includes only 2045+265, whose image separation, current estimate of
lensing galaxy redshift, and spectroscopic information on lensing galaxy type
are not consistent with current knowledge of galaxies (see Fassnacht et al.\
1999). For the multiply-imaged systems of Group~B, lensing galaxy types are 
unknown or uncertain. We do not exclude these systems but include them with a
penalization (see below).

However, the above prescription of excluding certain systems while keeping 
others is not a perfect solution. For example, proximate galaxies and/or 
nearby groups of galaxies outside the critical curves of single-lens systems
can have non-negligible effects. Indeed, current studies find that proximate
(satellite) galaxies (Cohn \& Kochanek 2004) and environmental groups 
(Keeton \& Zabludoff 2004) of the primary lensing galaxies 
can have various effects on the inferences on galaxies from
gravitational lenses.\footnote{The most significant effects of the lens 
galaxy environments appear to be biasing galaxy ellipticities and image
multiplicities.} However, the environmental effects on inferred velocity
dispersions appear to be relatively small (certainly smaller than statistical
errors arising from the current sample size of lenses).
Another concern may be that the primary galaxies of the systems of Group~D 
may be a biased sample of early-type galaxies (for example, preferentially 
more massive galaxies than average galaxies). Thus, excluding the systems of
Group~D could bias the image separation distribution of early-type `single' 
lenses. Cohn \& Kochanek (2004) argue that dropping compound lenses (i.e.\ 
lens systems with satellite galaxies) can be more biasing the image separation
distribution than simply including them as if they were single-galaxy lenses.
Notice, however, that Cohn \& Kochanek (2004) define a compound lens to be a 
system that contains galaxy(-ies) within 2 arcseconds of the primary galaxy. 
As a result, more than half of the Cohn \& Kochanek (2004) compound lenses do
not pertain to our Group~D. Therefore, the argument of Cohn \& Kochanek (2004)
does not directly apply to our analysis. Furthermore, as pointed out in \S 3.1
of Chae~(2003), the critical radii of the primary galaxies of multiple-galaxy
lenses (i.e.\ systems that contain multiple galaxies within their critical
radii) are on average similar to those of the single-lens galaxies, perhaps
implying that the primary galaxies of the multiple-lens systems may be a 
random sample of early-type galaxies. Nevertheless, we must bear in mind that
our analysis to follow potentially suffers from the above systematic biases;
taking them fully into account is beyond the scope of this work.

\begin{table*}
\caption{Multiply-imaged systems from the CLASS (including the JVAS; Browne 
et al.\ 2003) and the PANELS southern sky (Winn et al.\ 2000, 2001a,b, 
2002a,b) radio surveys. 
\label{complete}}
\begin{tabular}{clllllll}
\hline 
   &    &    & Source & Lens & Maximum & Number &Lensing \\
Group & Source & Survey  & redshift   & redshift 
 & image  & of & galaxy(-ies) \\
      &       &        & ($z_{s}$)  & ($z_{l}$) 
 & separation ($''$)  & images &  type \\
\hline
A & B0414+054  & JVAS  & 2.62   & 0.958 & 2.09   & 4  & early-type \\
A & B0445+123$^\dagger$  & CLASS &  ---   & 0.558 & 1.33   & 2 & early-type \\
A & B0631+519$^\dagger$  & CLASS &  ---   & 0.620 & 1.16   & 2 & early-type \\
A & B0712+472$^\dagger$  & CLASS &  1.34  & 0.41  & 1.27   & 4 & early-type \\
A & B1030+074  & JVAS  &  1.535 & 0.599 & 1.56   & 2  & early-type  \\
A & B1422+231$^\dagger$  & JVAS  &  3.62  & 0.34  & 1.28   & 4 & early-type \\
A & J1632$-$0033 & PANELS & 3.42  &  1$^*$  & 1.47   & 2  & early-type  \\
A & J1838$-$3427 & PANELS & 2.78  & 0.36$^*$ & 1.0   & 2  & early-type  \\  
A & B1933+503$^\dagger$  & CLASS &  2.62  & 0.755 & 1.17   & 4 & early-type \\
A & B1938+666  & JVAS  & $\ga 1.8$ & 0.881 & 0.93 & ring & early-type \\
A & B2319+051$^\dagger$  & CLASS & --- & 0.624/0.588 & 1.36 & 2 & early-type \\
\hline
B & B0128+437 & CLASS & 3.124 & 1.145$^*$ & 0.54 & 4 & unknown  \\
B & B0739+366 & CLASS & ---   &  ---    & 0.54   & 2  & unknown  \\
B & B1152+199$^\dagger$  & CLASS &  1.019 & 0.439 & 1.56   & 2  & unknown \\
B & B1555+375 & CLASS &  ---   &  ---    & 0.42   & 4  & unknown   \\
\hline
C & B0218+357$^\dagger$  & JVAS  &  0.96  & 0.68  & 0.334  & 2  & spiral \\
C & B0850+054$^\dagger$  & CLASS &  ---   & 0.588 & 0.68   & 2  & spiral \\
C & B1127+385 & CLASS &   ---  &  ---  & 0.70   & 2  & spiral  \\
C & B1600+434 & CLASS & 1.57   & 0.415 & 1.39   & 2  & spiral  \\
C & J2004$-$1349 & PANELS & --- &  ---  &  1.13  &  2  & spiral  \\
\hline
D & J0134$-$0931 & PANELS &  2.225 & 0.7645 & 0.681 & 2+4 & 2Gs  \\
D & B1359+154$^\dagger$  & CLASS &  3.235 & ---  & 1.65   & 6  &  3Gs \\
D & B1608+656$^\dagger$  & CLASS &  1.39  & 0.64  & 2.08   & 4  &  2Gs \\
D & B2108+213  & CLASS &  ---  & 0.365  & 4.55   & 2 or 3 & 2Gs+cluster \\
D & B2114+022$^\dagger$  & JVAS  & ---  & 0.32/0.59 & 2.57  & 2$^*$ & 2Gs \\
\hline
E & B2045+265$^\dagger$  & CLASS & ---   & 0.867  & 1.86   & 4  & puzzling \\
\hline
\end{tabular}

 $^\dagger$ Systems in the CLASS statistical sample (Chae 2003)

 $^*$ An estimated or assumed value

\end{table*}

We use the model of statistical lensing described in Chae (2003). We assume 
a spatially flat universe with an Einstein cosmological constant $\Lambda$ 
and adopt for the present matter density $\Omega_{\rm m} = 0.27$ (Spergel
et al.\ 2003), i.e.\ a concordance cosmological model favored 
by a broad range of astronomical observations 
(see, e.g., Chae et al.\ 2004 and references therein). We assume that
the distribution of early-type galaxies in luminosity is given by the 
Schechter (1976) form
\begin{equation}
\tilde{\phi}(L) dL=\tilde{\phi}_* \left(\frac{L}{L_*}\right)^{\tilde{\alpha}}
      \exp\left(-\frac{L}{L_*}\right) \frac{dL}{L_*}.
\end{equation}
Assuming a power-law relation between luminosity ($L$) and velocity 
dispersion ($\sigma$), i.e.
\begin{equation}
\frac{L}{L_*} = \left(\frac{\sigma}{\sigma_*}\right)^{\gamma},
\end{equation}
we can describe the distribution of early-type galaxies in velocity dispersion
in the following form\footnote{This particular parameterization was introduced
by Sheth et al.\ (2003).}
\begin{equation}
\phi(\sigma) d\sigma = \phi_* 
\left(\frac{\sigma}{\sigma_*}\right)^{\alpha} 
    \exp\left[-\left(\frac{\sigma}{\sigma_*}\right)^{\beta}\right] 
  \frac{\beta}{\Gamma(\alpha/\beta)} \frac{d\sigma}{\sigma},
\end{equation}
where we have the following relations: $\alpha = (\tilde{\alpha} + 1) \gamma$, 
$\beta = \gamma$, and $\phi_* = \tilde{\phi}_* \Gamma(\tilde{\alpha} + 1)$.

The particular differential probability that a source with redshift $z_s$ be
multiply-imaged with image separation $\Delta\theta$ by a distribution of 
galaxies at redshift $z_l$ following equation~(3) may be defined by
\begin{eqnarray}
\delta p & \equiv & \frac{d^2p}{dzd(\Delta\theta)}/\frac{dp}{dz} \nonumber \\
  & = & \frac{1}{2} \frac{\beta}{\Gamma[(\alpha+4)/\beta]} 
     \frac{1}{\Delta\theta_*}
  \left(\frac{\Delta\theta}{\Delta\theta_*}\right)^{\alpha/2+1}
  \exp\left[-\left(\frac{\Delta\theta}{\Delta\theta_*}\right)^{\beta/2}\right],
\end{eqnarray}  
where the differential probabilities $d^2p/dzd(\Delta\theta)$ and 
$dp/dz$ can be found in \S2.1.2 of Chae~(2003) and 
  the characteristic image separation $\Delta\theta_*$ is given by 
\begin{equation}
\Delta\theta_* = \lambda 8 \pi \frac{D(z,z_s)}{D(0,z_s)}
                 \left(\frac{\sigma_*}{c}\right)^2,
\end{equation}
where the lens potential is assumed to be that of a singular isothermal
ellipsoid (SIE), $\lambda$ is a dynamical normalization factor 
[see \S2.1.1 of Chae~(2003)], and $D(z_1,z_2)$ is the angular-diameter
distance between redshifts $z_1$ and $z_2$. For $\lambda$ we assume 
that early-type galaxies are not biased toward oblate or prolate shape
using the singular isothermal ellipsoid model of Chae~(2003); thus, 
$\lambda \approx 1$ [see {Fig.}~1 of Chae~(2003)].

For a sample of $N_{\rm{L}}$ multiply-imaged sources, the likelihood 
$\mathcal{L}$ of the observed image separations given the statistical lensing
model is defined by
\begin{equation}
\ln \mathcal{L} = \sum_{l=1}^{N_{\rm{L}}} w_l \ln \delta p_l,
\end{equation}
where $\delta p_l$ is the particular differential probability given by
equation~(4) for the $l$-th multiply-imaged source and $w_l$, a weight
factor, is set to unity for early-type lens systems (Group~A of Table~1)
and 0.8\footnote{Among the single-galaxy lenses with known galaxy types 
(Group A \& C) the fraction of early-type galaxies is 0.7, which is similar 
to the assumed value here.} for systems whose lensing galaxy types are unknown
(Group~B of Table~1). Using equation~(6), we define a ``$\chi^2$'' as follows:
\begin{equation}
\chi^2 = -2 \ln \mathcal{L}.
\end{equation}
We minimize the $\chi^2$ ({eq.}~7) to determine the maximum-likelihood 
values of model parameters of interest and obtain their confidence limits 
using the usual $\Delta\chi^2$ statistics. 

\section{Results and Discussion}

We first consider constraining the shape of velocity dispersion function 
(VF; {eq.}~3) and the characteristic velocity dispersion $\sigma_*$.
The adopted VF has then three effective parameters, 
i.e.\ $\alpha$, $\beta$, and $\sigma_*$.\footnote{Parameter 
$\phi_*$ has no relevance on the shape of the image separation distribution; 
it only plays the role of a multiplicative constant. This allows us to fix 
$\phi_*$ by an arbitrary constant here.} Because the lensing galaxies are 
distributed within $0.3 \la z \la 1.0$, these parameters to be constrained
may refer to an effective epoch of $z \sim 0.6$. Figure~1 shows confidence 
limits (CLs) on the parameter plane spanned by $\alpha$ and $\beta$. 
To obtain these CLs, parameter $\sigma_*$ has been varied to minimize 
the $\chi^2$ ({eq.}~7) at each grid point on the $\alpha$-$\beta$ plane. 
Figure~1 also shows the points corresponding to the recently measured
 SDSS local stellar VF (Sheth et al.\ 2003) and an inferred local stellar
VF (Chae 2003) based on the Second Southern Sky Redshift Survey (SSRS2) 
early-type luminosity function and a Faber-Jackson relation of the form
$L/L_* = (\sigma/\sigma_*)^{4.0\pm 0.2}$. Notice that these two stellar VFs
are significantly different from each other in their corresponding parameter
values.  From Figure~1 we find that (1) broad regions in the $\alpha$-$\beta$
plane are consistent with the present data and consequently, neither of the
parameters can be tightly constrained; (2) both the SDSS measured stellar VF
and the SSRS2 inferred stellar VF are consistent with the image separation 
distribution (lying nearly within the 68\% CL). 
Nevertheless, it appears that the shape of the SDSS VF is
marginally favored over that of the SSRS2 VF.

A by-product from Figure~1 is that parameter $\sigma_*$ is essentially
unconstrained by the data. This is because $\sigma_*$ varies as $\alpha$ and
$\beta$ are varied. What would be the (constrained) value of $\sigma_*$ if
$\alpha$ and $\beta$ are fixed by a stellar VF? Strictly speaking
 we need a stellar VF for the same epoch, i.e.\ $z \sim 0.6$. 
However, since a stellar VF at such a redshift is not available, 
we use local stellar VFs assuming non-evolution of the VF
from $z \sim 0.6$ to the present. Figure~2 shows CLs on 
$\sigma_*$ for the cases of fixing $\alpha$ and $\beta$ by the SDSS local 
central stellar VF and the SSRS2 local central stellar VF. In Figure~2 are also
shown the local central stellar values of $\sigma_*$ (Sheth et al.\ 2003; 
Chae 2003) for comparison. From Figure~2 we find that the lensing-based 
values of $\sigma_*$ are nearly equal to the corresponding stellar values for
the adopted stellar VFs. The ratio between the lensing-based SIE velocity 
dispersion and the spectroscopically-measured central stellar velocity 
dispersion is given by 
$f_{\rm SIE/center}(\equiv \sigma_{*{\rm SIE}}/\sigma_{*{\rm center}})
 =0.90\pm 0.18$ (SDSS) and $1.04\pm 0.19$ (SSRS2).
Treu \& Koopmans (2004) have studied five individual lens systems 
through both modeling the lens potentials using an SIE and optical 
spectroscopic observation of the lensing galaxies. Treu \& Koopmans (2004)
have determined the ratio of the SIE velocity dispersion to the stellar 
velocity dispersion for the five lens systems and found a mean value of 
$\langle \sigma_{\rm SIE}/\sigma_{\rm center} \rangle = 1.15\pm 0.05$.
The result by Treu \& Koopmans (2004) cannot be directly compared with the 
result of this work above. For the former the SIE and central stellar velocity 
dispersions were determined for the same objects but the objects do not 
well represent the population of early-type galaxies while for the latter
the SIE and central characteristic velocity dispersions were determined through
fitting to a model function for two different statistical samples that are
supposed to well represent early-type galaxy population. 
Nevertheless, our and Treu \& 
Koopmans (2004) results are in agreement.

What would be the appropriate interpretation of this agreement between the
lensing-based inferred velocity dispersion and the spectroscopically-measured
central stellar velocity dispersion? To answer this question we must first 
examine the relevance of the observed image separation to the adopted lens 
model (\S~2; Chae 2003). For the model of Chae~(2003) it is assumed that 
lensing is caused by a single-component model galaxy, i.e.\ a singular
isothermal ellipsoid (SIE). 
However, the real early-type galaxy that causes lensing
is of course not a single-component system; it can be divided into 
the luminous stellar component and the extended dark-matter halo component.
What is measured by the angular-splitting size of the 
gravitationally-lensed image is the total mass within
(the cylindrical region of) the Einstein radius of the lensing galaxy, 
which is of the order of the effective radius of the optical galaxy 
and (nearly) one half of the image separation. Results from 
detailed modeling of gravitationally-lensed systems 
(e.g.\ Rusin et al.\ 2004; Treu \& Koopmans 2004) 
and analyses of X-ray data of early-type galaxies (e.g.\ Loewenstein \&
White 1999) show that both stellar mass and dark mass are significant
contributors to the mass within the effective radius. 
The velocity dispersion distribution of a subsystem of the lensing
galaxy depends on both the total gravitational potential of the galaxy and the
mass profile of the subsystem. Therefore, the image separation probes a larger
scale compared with the spectroscopic technique and depends on the velocity
dispersions of both stars and dark matter.
This means that the velocity dispersion of an 
SIE constrained by the image separation cannot be simply regarded as the 
velocity dispersion of dark-matter particles (or, of stars). The SIE velocity
dispersion is just a theoretical parameter for the total mass
of stars and dark matter. Then, what would be the implication of the 
agreement between the two? In particular, would this agreement 
imply that dark matter velocity dispersion is equal to
stellar velocity dispersion? One simple possibility would be that
the mass profile is approximately isothermal and dark matter
and stellar velocity dispersions are equal. This interpretation would be in
agreement with the result by Kochanek (1994) that the dark matter halo and the
central stellar velocity dispersions are nearly equal based on modeling the 
line-of-sight velocity dispersion profiles of 37 elliptical galaxies using
a Jaffe stellar density profile and the SIS model for the total mass 
distribution. 
However, this would be in conflict with other independent results.
For example, combined analyses of X-ray data and spectroscopic data 
of elliptical galaxies (White \& Davis 1998; Loewenstein \& White 1999) 
find that dark-matter velocity dispersions are greater than stellar velocity 
dispersions ($\sigma_{\rm dm}^2 \approx 1.4-2\sigma_{\rm star}^2$; 
Loewenstein \& White 1999). Perhaps, this issue could be resolved 
in the future by incorporating a more realistic 
two-component galaxy model into the statistical lensing formalism. 

Until now the characteristic normalization (i.e.\ $\tilde{\phi}_*$ in {eq.}~[1]
 or $\phi_*$ in {eq.}~[2]) has been fixed because we did not use an absolute 
lensing probability but used only a relative lensing probability as a function
of image separation through equation~(4). However, it is interesting to ask
whether currently available characteristic normalizations of early-type
galaxies are consistent with currently available absolute lensing 
probabilities. At present, the most reliable sample for an absolute lensing 
probability is provided by the CLASS statistical sample
(see Table~1 and \S 3.1 of Chae 2003). 
For the present purpose we exclude the two spiral
lenses (0218+357, 0850+054) and the system 2045+265 for which there exist 
spectroscopic indications of late-type galaxy(-ies) (Fassnacht et al.\ 1999). 
We assume that the remaining 10 systems (0445+123, 0631+519, 0712+472, 
1152+199, 1359+154, 1422+231, 1608+656, 1933+503, 2114+022, 2319+051) are all
early-type lenses. However, we ignore the image separations of the three 
systems 1359+154, 1608+656, and 2114+022 whose lenses are multiple galaxies;
by this we mean that we do not use their image separations to constrain the
velocity dispersion function. We then use the same likelihood analysis method
as used in Chae~(2003). Notice here particularly that the above multiple-galaxy
lenses are included in lensing rate with the adjustment that $\sigma_*$ is 
replaced by $n^{1/4} \sigma_*$ in the lensing probability for a multiple-galaxy
lens system with $n$ galaxies.\footnote{Since the multiple-imaging probability
by a population of single galaxies is proportional to $\sigma_*^4$, the 
probability for lensing by $n$ galaxies may be assumed to be proportional to
$n \sigma_*^4$ if ignoring the interference of the combined potentials.} 
Since not all four parameters (i.e.\ $\alpha$, $\beta$, $\sigma_*$, and 
 $\tilde{\phi}_*$\footnote{Here this parameter is used rather than $\phi_*$
because the latter diverges for $\alpha = 0$ which we want to consider.}) can
be constrained even when an absolute lensing probability is included, here we
fix parameters $\alpha$ and $\beta$ using the SSRS2 or SDSS stellar VFs and
constrain the parameter plane spanned by $\sigma_*$ and $\tilde{\phi}_*$.

Figure~3 shows the confidence limits in the parameter plane spanned by 
$\sigma_*$ and $\tilde{\phi}_*$ for the cases of fixing $\alpha$ and $\beta$
using the SDSS or SSRS2 local stellar VFs. Here the values of $\sigma_*$ and
$\tilde{\phi}_*$ from the stellar VFs (Chae~2003; Mitchell et al.\ 2005) are
also marked and compared with the likelihood regions. The interesting points 
are the following. First, the best-fitting point for the SDSS VF is slightly 
a better overall fit than that for the SSRS2 VF comparing the maximized values
of the likelihood for the two cases (specifically, $\Delta \chi^2 \la 1$). 
This is consistent with the results shown in Figure~(1). Second, while the 
value of $\tilde{\phi}_*$ from the SSRS2 VF lies in the corresponding most 
likely region, that from the SDSS VF does not. Specifically, the SDSS measured 
$\tilde{\phi}_*$ is a factor of $\sim 3$ lower than the best-fitting value 
based on the CLASS absolute lensing probability. 
This suggests that the SDSS selection process of early-type galaxies 
(Bernardi et al.\ 2003; Mitchell et al.\ 2005)
has significantly underestimated the abundance of morphologically early-type 
galaxies. Indeed, the SDSS selection process (Bernardi et al.\ 2003; Mitchell
et al.\ 2005) excludes the galaxies that have centrally-concentrated light 
distributions but have spectra showing recent star-formation activities.
However, as far as lensing is concerned, such a galaxy acts as an early-type
because its centrally concentrated light distribution makes it have a 
relatively large multiple-imaging cross section. The lesson is that
morphologically and dynamically early-type galaxies may show non-traditional
colors and spectra. A re-classification of SDSS galaxies based on a new method
that agrees relatively well with visual classifications shows that early-type 
galaxies are much more abundant than the Bernardi et al.\ (2003) and the
Mitchell et al.\ (2005) estimates, probably in good agreement with the 
the absolute lensing probability of the CLASS statistical sample as
analyzed in this work (C. Park, et al., private communications). 
 Therefore, we argue that
Mitchell et al.\ (2005) may have overestimated $\Omega_\Lambda$ owing to
their use of an underestimated $\tilde{\phi}_*$.\footnote{Notice that
Mitchell et al.\ (2005) used for calculating magnification bias an index 
$\eta = 2.07$ in the differential number-flux density relation 
$|dN/dS_{\nu}| \propto S_{\nu}^{-\eta}$ which is steeper than the Chae et al.\
(2002) and Chae (2003) adopted value of $\eta = 1.97$. This helped lower their
estimated value of $\Omega_{\Lambda}$; otherwise they would have obtained an
even larger $\Omega_{\Lambda}$.} 

In conclusion, based on the statistics of multiple-imaging in the largest
uniformly-selected sample of gravitationally-lensed systems and adopting a SIE
 model for galaxies, we find that: (1) the lensing-based
SIE velocity dispersion agrees with the central stellar velocity dispersion
[$f_{\rm SIE/center} = 0.90\pm 0.18 (1.04 \pm 0.19)$ assuming the shape of
 SDSS (SSRS2) VF], (2) the shape of the SDSS stellar VF is in excellent 
agreement with the image separation distribution of multiply-imaged
systems while that of the SSRS2 stellar VF is also consistent, 
and (3) the abundance of morphologically early-type galaxies implied
by the CLASS absolute lensing probability under the current concordance
cosmological model is in good agreement with that of the SSRS2 early-type
galaxies but significantly higher than that of the SDSS early-type galaxies.

\bigskip

KHC acknowledges support from the ARCSEC of KOSEF.
We thank S. Mao, M. Park, and R. Sheth for comments and/or discussions.
We also thank the observers (of the CLASS sources) whose hard-work results
have made this analysis possible. The anonymous referee provided a critical
reading of the manuscript that helped improve the presentation and discussion.


\begin{figure}
\centerline{\epsfig{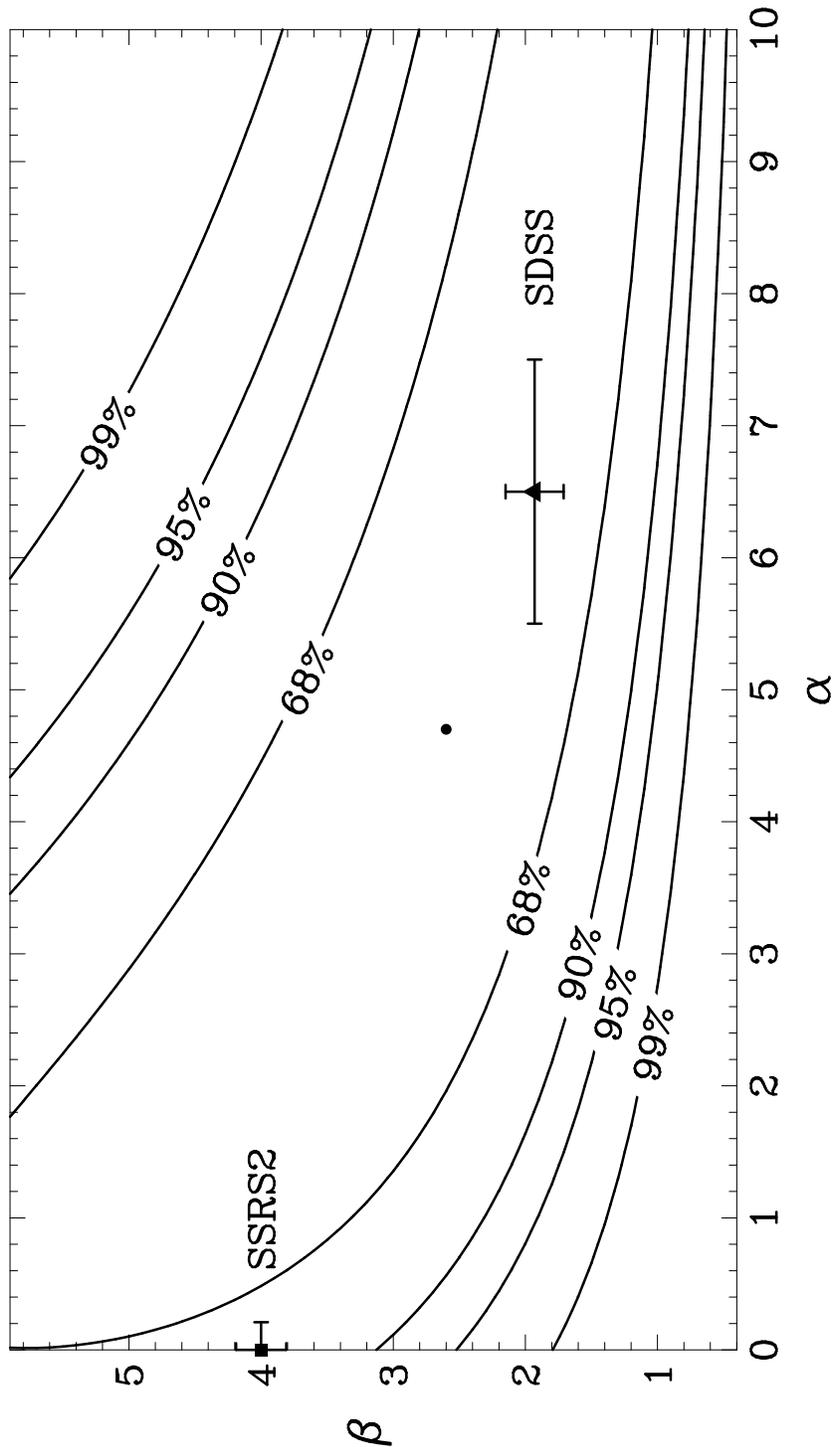}}
\caption{Confidence limits of 68, 90, 95, and 99\% on the shape of the model
velocity dispersion function of early-type galaxies (equation~3) based on a 
statistical sample of gravitationally lensed image separations and the lensing
formalism of Chae (2003). The measured/inferred shapes of the SDSS and SSRS2 
central stellar velocity dispersion functions are also shown for comparison.}
\label{f1}
\end{figure} 

\begin{figure}
\centerline{\epsfig{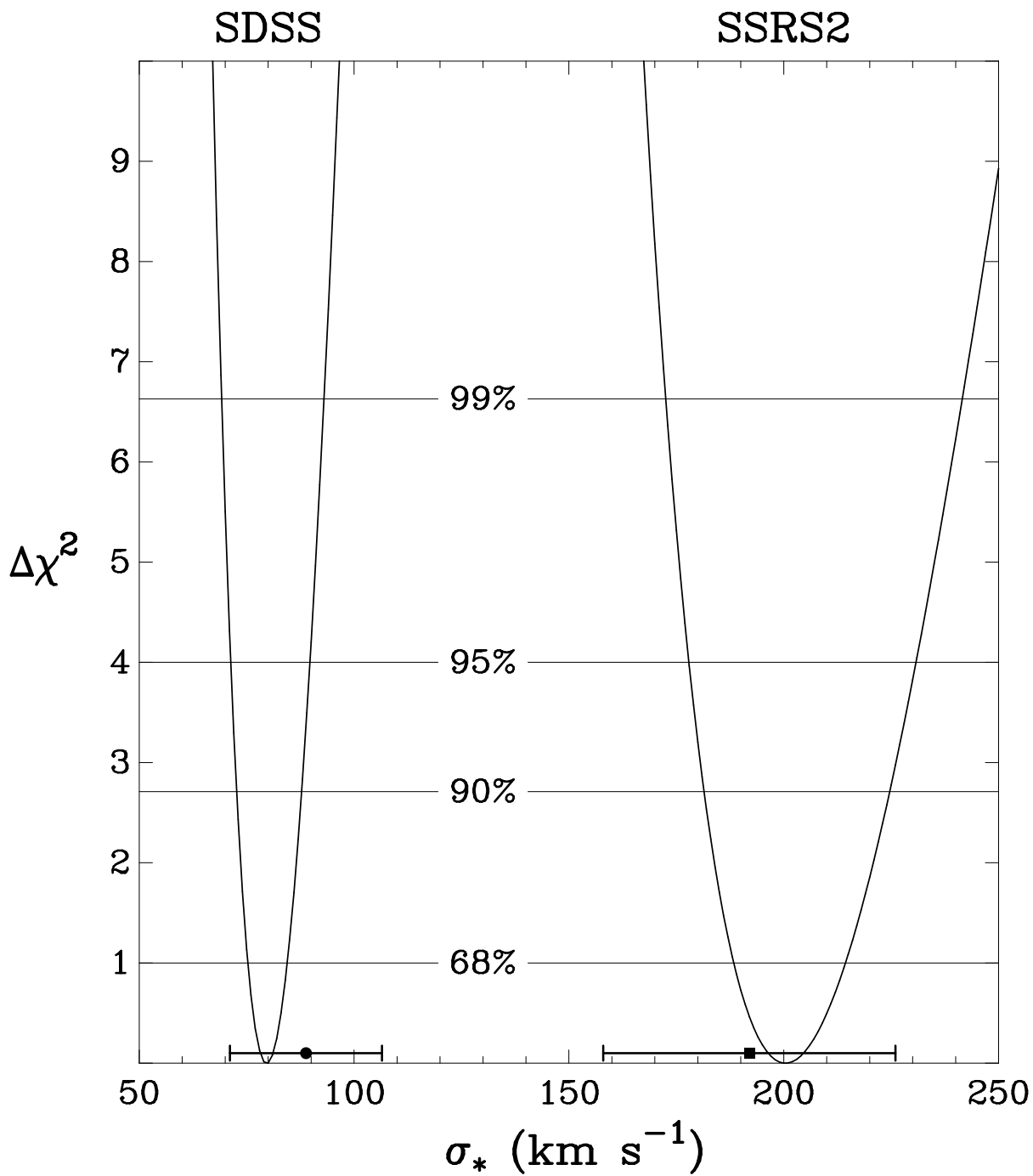}}
\caption{Confidence limits of 68, 90, 95, and 99\% on the lensing-based 
characteristic velocity dispersion ($\sigma_{*{\rm SIE}}$) of early-type 
galaxies. For this the shape of the velocity dispersion function (equation~3)
is fixed by either the SDSS or SSRS2 central stellar velocity dispersion 
function. The lensing-based characteristic velocity dispersions are 
respectively compared with the SDSS measured and SSRS2 inferred characteristic
velocity dispersions of stars of the central regions ($\sigma_{*{\rm center}}$)
of early-type galaxies.}
\label{f2}
\end{figure} 

\begin{figure}
\centerline{\epsfig{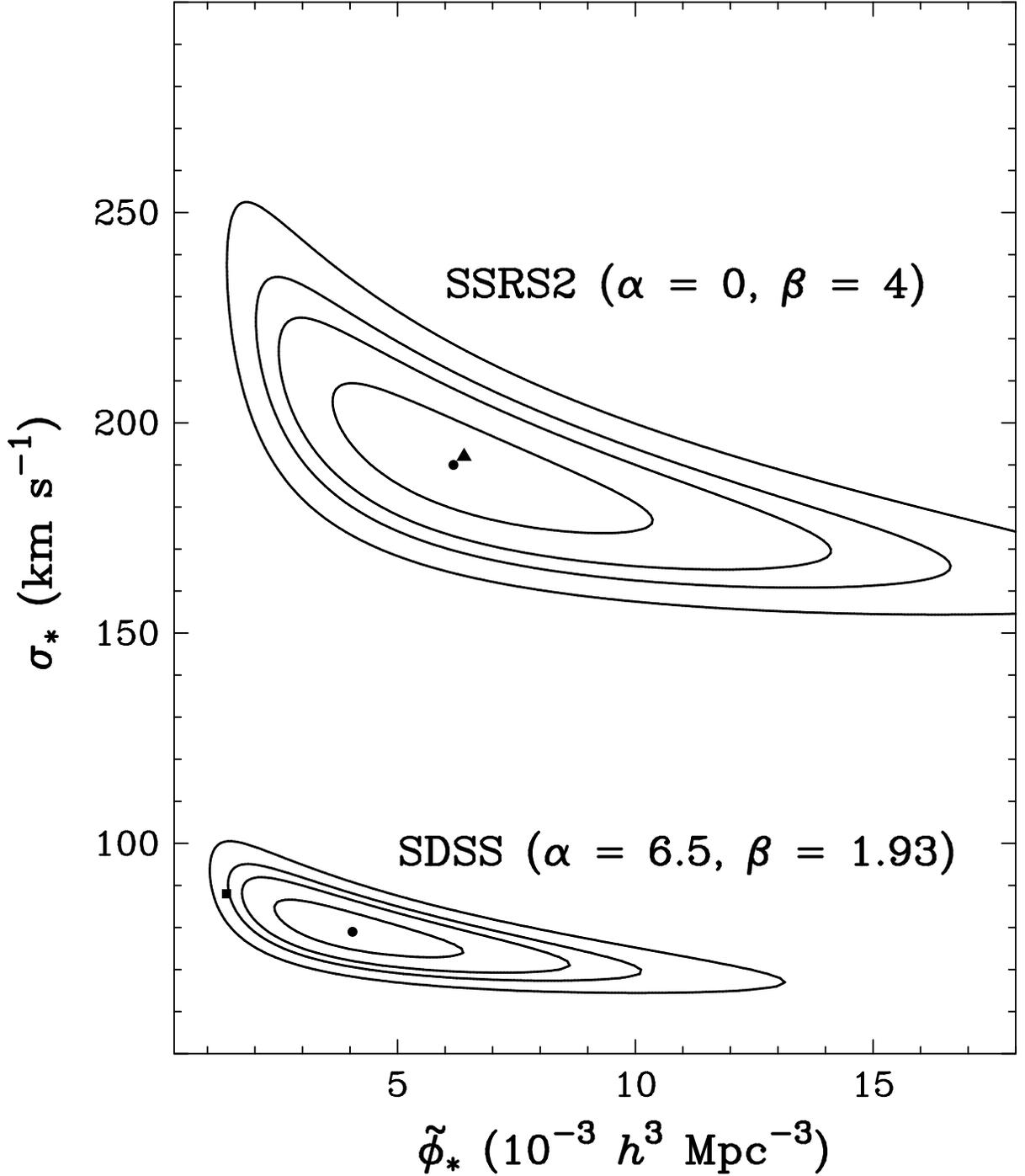}}
\caption{Confidence limits of 68, 90, 95, and 99\% on the plane spanned by
the normalization $\tilde{\phi}_*$ and the characteristic velocity dispersion
$\sigma_*$ (eq.~1) for the fixed shapes of the velocity dispersion function.
Here the SDSS and SSRS2 local central stellar velocity dispersion functions
of early-type galaxies are considered. These constraints are based on the
statistics of the CLASS statistical sample and the same analysis method as 
used in Chae~(2003). The points marked by square and triangle represent 
respectively the normalizations and characteristic velocity dispersions
from the SDSS and SSRS2 stellar VFs.}
\label{f3}
\end{figure}

\end{document}